%% file: stability_scores.arxiv.tex
\documentclass[10pt,twocolumn]{article}
\usepackage{stmaryrd}
\usepackage{amsmath}
\usepackage{amssymb}
\usepackage{amsthm}
\usepackage{times}
\usepackage{algpseudocode}
\usepackage{multirow}
\usepackage{slashbox}
\usepackage{enumerate}
\usepackage[hypertex]{hyperref}

\newtheorem{theorem}{Theorem}
\newtheorem{corollary}[theorem]{Corollary}
\newtheorem{lemma}[theorem]{Lemma}

\newtheorem{proposition}[theorem]{Proposition}

\def\({\left(}
\def\){\right)}
\DeclareMathOperator*{\avg}{avg}
\DeclareMathOperator*{\argmin}{argmin}

\newcommand{\labeq}[2]{   
\begin{equation}
\label{eq:#1}
#2
\end{equation}}
\newcommand{\tup}[1]{\left\langle #1\right\rangle}
\renewcommand{\vec}[1]{\mathbf{#1}}
\def\mc{\mathcal}
\def\({\left(}
\def\){\right)}
\newcommand{\floor}[1]{\left\lfloor #1 \right\rfloor}
\newcommand{\ceil}[1]{\left\lceil #1 \right\rceil}
\def\eps{\epsilon}

\newenvironment{rtheorem}[1]{\medskip\noindent\sc{Theorem~\ref{#1}.}\begin{itshape}}{\end{itshape}}
\newenvironment{rprop}[1]{\medskip\noindent\sc{Proposition~\ref{#1}.}\begin{itshape}}{\end{itshape}}
\newenvironment{rlemma}[1]{\medskip\noindent\sc{Lemma~\ref{#1}.}\begin{itshape}}{\end{itshape}}

\newcommand\newpar[1]{\vspace{-0mm}\subsubsection*{#1}}
\newcommand\subpar[1]{\medskip\vspace{-0mm}\noindent \textbf{#1.} }

\allowdisplaybreaks
\begin{document}
\def\emailsize{\large}
\def\affsize{\normalsize}
\title{Stability Scores: Measuring Coalitional Stability}

\author{Michal Feldman\thanks{Microsoft Israel R\&D Center and Hebrew University of Jerusalem}
\and
Reshef Meir\thanks{Microsoft Israel R\&D Center and Hebrew University of Jerusalem}
\and
Moshe Tennenholtz\thanks{Microsoft Israel R\&D Center and Technion-Israel Institute of Technology}\\
}

\maketitle

\begin{abstract}
We introduce a measure for the level of stability against coalitional deviations, called \emph{stability scores}, which generalizes widely used notions of stability in non-cooperative games.
We use the proposed measure to compare various Nash equilibria in congestion games, and to quantify the effect of game parameters on coalitional stability.
For our main results, we apply stability scores to analyze and compare the Generalized Second Price (GSP) and Vickrey-Clarke-Groves (VCG) ad auctions.
We show that while a central result of the ad auctions literature is that the GSP and VCG auctions implement the same outcome in one of the equilibria of GSP, the GSP outcome is far more stable. Finally, a modified version of VCG is introduced, which is group strategy-proof, and thereby achieves the highest possible stability score.
\end{abstract}
%
%
%

\section{Introduction}
\label{SEC:intro}
One of the most basic questions of game theory is: given a game in strategic form, what is its solution?
By \emph{solution} we typically mean a strategy profile that can be proposed to all agents, and no rational agent would want to deviate from it. Thus a solution should be \emph{stable}.
Many solution concepts for games have been studied; these studies differ by the level and interpretation of stability, as well as by the underlying assumptions that are required to achieve it.
The best known solution concept for games is the Nash equilibrium (NE), a strategy profile from which no agent has an incentive to deviate \emph{unilaterally}.

A basic problem with the NE solution concept is
that a NE does not take into account joint deviations by coalitions of players.
We usually assume that an individual will deviate from a profile if she has an available strategy that strictly
increases her payoff.
In some settings it would be natural to assume also that a group of individuals will deviate if they have an available joint strategy that strictly increases the payoff of each group member.
The \emph{Strong Equilibrium} (SE) concept by Aumann~\cite{Aumann59} deals with this problem.
A  profile is a SE if no coalition of agents can jointly deviate in a way that strictly increases the payoff of each coalition member. Intermediate levels of coalitional stability have been suggested, such as stability against deviations of small coalitions (see e.g. \cite{AndelmanFeldman}), and in particular pairs.
An even more appealing solution concept than SE is the \emph{Super-Strong Equilibrium} (SSE) that considers deviations in which no member loses while at least one member makes a positive gain (see, for example, \cite{FelTen09}).

A major problem with these proposed solutions is that they seldom exist.
Indeed, SSE rarely exist even in cases where strong equilibria do exist (e.g., in simple congestion games \cite{HolzmanCong1,AndelmanFeldman}), and even if only deviations by pairs are considered.

In this paper we relax the strong requirement that no coalition will have an incentive to deviate, and suggest a quantitative measure to coalitional stability.
Assuming we have a Nash equilibrium profile of a game where some pairs of agents can still deviate, we may still wish to measure its stability by referring to the {\em number} of pairs that have beneficial deviations from that profile.
More generally, given a game and a strategy profile, we can associate with it a tuple in which the $r$-th entry in the tuple is the number of coalitions of size $r$ that can gain by a deviation.
This tuple determines the {\em stability score} of the  strategy profile.

Given two strategy profiles, we need a way to decide which one is more stable.
A common practice in game theory is to prefer strategy profiles that are \emph{in equilibrium}, i.e. in which there are no unilateral deviations.
Since small coalitions are more likely to form and maintain cooperation, a natural extension is to compare stability scores of games with associated strategy profiles using a \emph{lexicographic} ordering of the corresponding vectors.\footnote{There are many ways to compare stability score vectors. Choosing the ``right'' one highly depends on the context and underlying assumptions. However in this paper we avoid such complications by only comparing deviations of coalitions of the same size.}
For example, given two $n$-person games, $G_1$ and $G_2$, with respective Nash equilibria $s_1$ and $s_2$, the stability score of the former will be higher if the number of beneficial deviations by pairs from $s_1$ in $G_1$ is smaller than the number of beneficial deviations by pairs from $s_2$ in $G_2$. 

While the existence of, say, $19$ coalitions that can deviate rather than $15$ does not have much significance, we usually care about the behavior in some parameterized family of games where parameters may include number of players, size of the strategy space, etc.  If the score of $\vec a$ is \emph{asymptotically lower} than the score of $\vec  a'$ (w.r.t. one of the parameters), then this may indicate that $\vec a'$ is substantially more prone to coalitional deviations.

Moreover, when studying such a parametrized family, stability scores may assist us in understanding how the parameters of the game affect coalitional stability. This holds even if there is a unique or a prominent equilibrium. 

Stability scores are particularly useful in the context of \emph{mechanism design}, as they allow us to quantify the coalitional stability of various mechanisms and to compare mechanisms that operate in a specific domain.
To illustrate this point, we consider two central mechanisms in what is perhaps the most widely studied economic setup in recent years: ad auctions.
We analyze in detail the Generalized Second Price (GSP) auction and the Vickrey-Clarke-Groves (VCG) auction, and compare their stability scores.

\subsection{Related work}

\newpar{Related solution concepts in games}

In the context of non-cooperative games approximate stability is typically measured by the strength of the incentive required to convince an agent to deviate, captured for example by the concept of $\epsilon$-Nash equilibrium. As discussed above, stability against collusion is captured by concepts such as SE and SSE,  but these often do not allow a fine distinction between various outcomes.

In addition, coalitions are the key component in \emph{cooperative} game theory, and many variations of coalitional stability have been studied. While we are unaware of solutions concepts that quantify stability by measuring coalitional deviations, models of restricted cooperation capture social constraints that may prevent the formation of some coalitions~\cite{Myerson77}. Thus a (cooperative) game may not be stable against every coalitional deviation (i.e. have an empty core), but still satisfy all the coalitions that can form in practice. Recently, some papers studied how such social context affects the stability of the game~\cite{Demange04,MRM:2011:IJCAI}. Moreover, even if some coalition \emph{can} gain by deviation, it may or may not do so: Members of the coalition might intentionally avoid cooperation based on far-sighted prediction (an assumption underlying \emph{coalition-proofness} for example~\cite{CProof87}), or just fail to recognize the benefit in deviating. This is especially true if the coalition is large.  Stability scores do not assume a particular social context or incentive structure, but simply try and minimize the number of  coalitions with profitable deviations. 

\newpar{Collusion and equilibria in ad auctions}
Major results of previous work on ad auctions, characterized a special family of equilibria of GSP the auction (used in practice), termed \emph{Symmetric Nash Equilibria}, or SNE (see Section~\ref{sec:auction_model} for details)~\cite{Varian07}. SNEs have many attractive properties which make them a natural choice as outcomes of the GSP auction. Moreover, it has been shown that the SNE leading to the lowest revenue for the seller (termed \emph{Lower Equilibrium} (LE)), coincides with the natural equilibrium of VCG where all bidders report their true values.

The above results led to a surge of papers comparing VCG and the various equilibrium outcomes of GSP, under both public information and private information settings~\cite{Kuminov09,TLB09,ES2010,LPT11}. However, these comparisons focused mainly on revenue, rather than on coalitional stability.
The VCG mechanism was shown to be vulnerable to collusion in various domains (see, e.g., ~\cite{ConSan06,Bachrach10} for relatively recent work), compared to a simple first-price (pay-your-bid) auction.
The formal literature on collusion in second-price auctions goes back to Graham and Marshall \cite{GrahamMarshall}, while the literature on the more involved matter of collusion in first-price auctions  goes back to McAfree and McMillan \cite{McAfeeMcMillan}.

\subsection{Our contribution}
Stability scores are formally defined in Section~\ref{SEC:prem}, where we show how they generalize well known solution concepts. In Section~\ref{SEC:RSG} we study strict stability scores in a simple family of congestion games. The main purpose of this study is to demonstrate how stability scores can be used in order to compare different Nash equilibria, and to measure how stability is affected by game's parameters. Moreover, while the studied family itself is quite simple, it is often used to model real world situations such as load balancing. Our analysis can give some intuition as to the main factors affecting coalitional stability in such games. 

The main results are in Section~\ref{SEC:AUCTIONS}, where we present the VCG and GSP mechanisms for ad auctions (adopting the original model advocated for that setting in the seminal work by Varian~\cite{Varian07} and by Edelman et al.~\cite{Edelman_Ostrovsky_Schwarz_07}), and show bounds on stability scores in these auctions. 
In particular, we study how the stability of GSP varies as a function of the distributions of agents' valuations and slots' click-through rates, thereby showing that under certain reasonable conditions GSP is far more stable than VCG.

In Section~\ref{SEC:RESERVE} we introduce a modification to the VCG auction that can be used to overcome the observed instability of VCG.
In particular, we show that a revised VCG, in which a random reserve price is introduced, induces truth-telling as a super-strong equilibrium.

 Most proofs are deferred to the appendix to allow continuous reading.

\section{Preliminaries}
\label{SEC:prem}

\newpar{Games and equilibria}
Let $G=\tup{N,\{A_i\}_{i\in N},\{u_i\}_{i\in N}}$ be a normal form game, where $N=\{1, \ldots, n\}$ is the set of players, $A_i$ is the set of actions available to player $i$, and $u_i: \vec  A \rightarrow \mathbb R$ is player $i$'s utility, where
$\vec A=A_1\times\cdots\times A_n$ is the set of joint actions (profiles), and for every $\vec a\in \vec A$, 
$u_i(\vec a)$ denotes the utility of player $i$ under action profile $\vec a$. 
The vector of actions of all players except player $i$ in the profile $\vec a$ is denoted by $a_{-i}$.
An action profile $\vec a \in \vec A$ is a \emph{Nash Equilibrium} (NE) if $u_i(\vec a) \geq u_i(b_i,a_{-i})$ for every agent $i\in N$ and every alternative action $b_i \in A_i$.

When considering coalitions, given an action profile $\vec a$, we denote by $a_S$ the profile of agents in $S$, and by $A_S$ the set of all such joint actions. The  profile of all agents in $N \setminus S$ is denoted by $a_{-S}$.

Given a profile of actions $\vec a \in \vec A$, $b_S\in A_S$ is a \emph{strict deviation} from $\vec a$ if $u_i(b_S,a_{-S}) > u_i(a_S,a_{-S})$ for every $i \in S$. The profile $\vec a$ is termed a {\em Strong Equilibrium} (SE) if there are no ${S\subseteq N}$ and $b_S\in A_S$, such that $b_S$ is a strict deviation from $\vec a$.

One can also consider the following weaker notion of deviation.
Given a profile of actions $\vec a \in A$, $b_S\in A_S$ is a \emph{deviation} 
from $\vec a$ if $u_i(b_S,a_{-S}) \geq u_i(a_S,a_{-S})$ for every $i \in S$ and there exists $j \in S$ such that $u_j(b_S,a_{-S}) > u_j(a_S,a_{-S})$ . The profile $\vec a$ is termed a {\em Super-Strong Equilibrium} (SSE) if there is no ${S\subseteq N}, b_S\in A_S$ that is a deviation from $\vec a$. Since every strict deviation is clearly a deviation, every SSE is also a SE.

SSE captures the natural requirement that we should resist even situations in which a deviation only benefits some of the deviators without hurting others. A strategy profile is $r$-SE (respectively, $r$-SSE) if there are no coalitions of size at most $r$ that have strict deviations (resp., deviations).

\newpar{Stability scores}
The stability score of the profile $\vec a$ in game $G$ is defined as a vector with $n$ entries.
For every $1 \leq r\leq n$, let $\mc D_r(G,\vec a) \in \mathbb N$ (respectively, $\mc{SD}_r(G,\vec a) \in \mathbb N$) be the \emph{number} of coalitions of size $r$ that have deviations (resp., strict deviations) from $\vec a$ in $G$.
While there are many ways to impose an order on equilibria based on these vectors, we believe that the following lexicographic order is particularly natural.

Given two $n$-player games $G$ and $G'$ and two profiles $\vec a$ and $\vec a'$ in the respective games, we say that the pair $(G,\vec a)$ is \emph{more resistant to deviations} (or \emph{more stable})  than $(G',\vec a')$, if there exists some $r\leq n$ such that $\mc D_r(G,\vec a) < \mc D_r(G',\vec a')$ and the terms are equal for every $r'<r$.
We can similarly compare strict stability scores to one another. 

Our definition of stability score generalizes some widely used notions of stability. For example, $\vec a$ is a Nash equilibrium (NE) of $G$ iff $\mc D_1(G,\vec a)=\mc{SD}_1(G,\vec a)=0$. 
This means that the score of a NE (by either definition) is  always strictly better than the score of any profile that is not a NE. Further, any profile that is $r$-SE has a better strict-stability score than any non $r$-SE profile. A similar property holds w.r.t. $r$-SSE.
As a different example, a profile $\vec a$ is \emph{Pareto efficient} in $G$ iff $\mc{D}_n(G,\vec a) = 0$.


\section{Resource Selection Games}
\label{SEC:RSG}
In this section we demonstrate how stability scores can be used to measure and compare the stability of different outcomes in a given game. To this end we focus on a very simple parametrized family, where games are known to posses at least one pure equilibrium. A natural choice is the family of \emph{resource selection games}  (RSG) with identical resources.

In a RSG there is a set of resources $F=\{1,\ldots,m\}$, and a non-decreasing cost function $c:[n]\rightarrow \mathbb R_+$, where $[n]=\{1,\ldots,n\}$. Each agent $i\in N$ can select exactly one resource $j$, and suffers a cost (negative utility) of $c(n_j)$, where $n_j$ is the number of agents that selected resource $j$. RSGs are \emph{potential games} and thus always admit a pure Nash equilibrium. In fact, any NE $\vec a$ of a RSG $G=\tup{F,N,c}$ is a \emph{strong equilibrium}~\cite{HolzmanCong1}, and thus all equilibria have the same (strict) stability score. However, this is no longer true if the games are concatenated in a sequence.

Formally, a \emph{sequential} RSG (SRSG) is a RSG with $k$ steps. Thus a strategy of an agent $a_i\in F^k$ requires selecting one resource in each step (actions may not depend on the previous steps).\footnote{Equivalently, the game can be described as a routing game, with $k$ sequential parts and $m$ parallel edges in each part.} We next show that the number of coalitional deviations significantly depends on the played equilibrium. We consider games where  $m,n,k\geq 2$, focusing mainly on games with $2$ steps.

\subsection{Counting deviations: an example}
Suppose that $m=4,n=6,k=2$ and that $c(t)=t$ for all $t\leq n$. Any profile in which there are exactly 1 or 2 agents on each resource (in each step) is a Nash equilibrium. However, these equilibria differ in their stability against strict deviation of pairs. Suppose that in the first step agents are partitioned $\{1,2\},\{3,4\},$ $\{5\},\{6\}$, and repeat the same actions in the second step. Denote this profile by $\vec a$. In this case the pair $\{1,2\}$ can strictly gain as follows: agent~$1$ joins agent~$5$ (or $6$) in the first step, and agent~$2$ joins $5$ in the second. Thus the cost for each of the two agents drops from $4$ to $3$. The pair $\{3,4\}$ can do the same, thus $\mc{SD}_2(G,\vec a) = 2$.

On the other hand, consider a profile $\vec b$ where players play in the first step as in $\vec a$, and in the second step are partitioned $\{1,3\}, \{2,4\},$ $ \{5\},\{6\}$; then no pair can strictly gain by deviating. Notice though, that this is still not a strong equilibrium, as the coalition $\{1,2,3,4\}$ can still gain (agents $2,3$ deviate in the first step, and $1,4$ in the second), thus $\mc{SD}_2(G,\vec b)=0$ and $\mc{SD}_4(G,\vec b) = 1$.

Finally, in profile $\vec c$ agents are partitioned $\{1,5\},\{2,6\},\{3\},\{4\}$ (in the second step), and this is a strong equilibrium, i.e. $\mc{SD}_r(G,\vec c)=0$ for all $r$.  It therefore follows that w.r.t strict stability scores $\vec c$ is \emph{more stable} than $\vec b$, which is more stable than $\vec a$.

Note however that none of these profiles is an SSE or even $2$-SSE. More generally, in \emph{any} profile in $G$ there is at least one pair (in fact two) that shares a resource and thus they have a (weak) deviation where just one of them gains. Thus for every profile $\vec p$ in $G$, we have that $\mc{D}_2(G,\vec p) \geq 2$.

\subsection{Bounding stability scores in two-step RSG}
The example above shows that different NE profiles in a particular game may differ in their stability to deviations of pairs or larger coalitions. We want to get a better picture of the gap between the most and least stable NE profiles, focusing on pair deviations. For the results in this section, we will restrict our cost function to be convex.

A nondecreasing cost  function $c :[n] \rightarrow \mathbb R$ is said to be \emph{convex} if it has an increasing marginal loss; i.e., $c(i+1) - c(i) \leq c(j+1)-c(j)$ for every $i<j$. Note that when facing a convex cost function, agents in an RSG try to minimize the maximal number of agents using a single resource. If the number of agents on every resource is the same, we say that the partition is \emph{balanced}. If these numbers differ by at most one, we say that the partition is \emph{nearly balanced}. 

Let $G$ be a two-step game with a convex cost function. Note that when $n \mod m=0$, any NE is a balanced partition of agents to resources (in each step). In such partition, no coalition can gain by deviating, as at least one deviating agent will end up paying more in expectation. If, in addition, costs are \emph{strictly} convex, then even weak deviations are impossible. Since in this setting every NE is an SE (and even an SSE), stability scores are trivial. We therefore assume that $n\mod m = q> 0$.

 Let $\hat{\vec a}$ be the profile with the highest number of pair deviations, and let $\vec a^*$ be the profile with the lowest number of pair deviations. 

\begin{proposition}$\mc{SD}_2(G,\hat{\vec a})=\Theta\(\frac{qn^2}{m^2}\)$.
\label{th:cong_two_hat_a}
\end{proposition}
\begin{proof} [sketch of lower bound]
We note that in $\hat{\vec a}$ agents play some nearly balanced partition in the first step, and repeat the same partition in the second step. Thus some resources (called \emph{full}) will have $\ceil{n/m}$ agents, and the others will have $\floor{n/m}$ agents.
A crucial observation used in the proof (and in the proofs of the other propositions in this section), is  that a pair has a strict deviation if and only if it shares a full resource in both steps. Then (similarly to the example above) one agent switches to a non-full resource in the first step, and the other does the same in the second step.
%
\end{proof}

Note that when $q=\Theta(m)$, which is a typical situation, there are over $\Omega\(\frac{n^2}{m}\)$ deviating pairs.

We find that the best NE $\vec a^*$ is significantly better than $\hat{\vec a}$.

\begin{proposition}\label{th:cong_two_a_star}
$\mc{SD}_2(G,\vec a^*)=O\( \frac{n^2 }{  \rule{0pt}{1.8ex} m^2}\)$. Further, if either $n<m^2$ or $q\leq \frac{m}{2}$, then $\mc{SD}_2(G,\vec a^*)=0$, i.e. $\vec a^*$ is $2$-SE.
\end{proposition}
In order to achieve the upper bound asserted in the proposition we define a profile that tries to scatter in the second step  agents that shared a resource in the first step.
 As a qualitative conclusion, we see that in order to minimize possible deviations, agents should form a partition in the second step that differs as much as possible from the partition in the first step.
\subsection{SRSGs with many steps}

The following proposition quantifies the stability score of a random pure NE in a RSG with $k$ steps. Note that the set of pure NEs coincides with the set of profiles that are nearly balanced in each step.

\begin{proposition}\label{th:cong_random}   
Let $G$ be an SRSG with $k$ steps and a convex cost function, and let $\vec a$ be a random NE in $G$. The expected number of deviating pairs in $G$ is $\mc{SD}_2(G,\vec a) \cong \binom{n}{2}\(1-(1+\alpha)e^{-\alpha}\)$, where $\alpha=\frac{q(k-1)}{m^2}$.
\end{proposition}
We can summarize how the parameters affect stability as follows. If the number of steps $k$ is small, and the number of resources $m$ increases, then $\alpha\rightarrow 0$, and thus $\mc{SD}_2(G,\vec a)\rightarrow 0$ as well (i.e. there are very few pairs that can deviate). Conversely,  when the number of steps grows (in particular when $k\gg\frac{m^2}{q}$), then almost every pair can deviate with a high probability.

As a corollary of Proposition~\ref{th:cong_random} when $k=2$, we get the lower bound of Proposition~\ref{th:cong_two_hat_a} for the case $q=\Theta(m)$, as 
\begin{align*}
\mc{SD}_2(G,\hat{\vec a}) &\geq \binom{n}{2} \(1-\(1-\frac{1}{m}\)\(1+\frac{1}{m}\)\)
= \Omega\(\frac{n^2}{m}\).
\end{align*}

\section{Stability Scores in Ad Auctions}
\label{SEC:AUCTIONS}
Having showed how stability scores can be used to analyze coalitional stability in simple games, we next turn to prove our main results. We compute the stability scores of the  VCG and GSP ad auctions, which  are central to the recent literature on economic mechanism design. Since both auctions admit strong equilibria, we do not consider strict deviations, and instead focus our analysis on weak deviations and the scores they induce. 
\subsection{Ad auctions: model and notations}
\label{sec:auction_model}
An ad auction has $s$ slots to allocate, and $n\geq 2s$ bidders,\footnote{When discussing deviating pairs it is sufficient to assume $n>s$, which is a typical situation. Also, all of our results can be easily adjusted to cases with fewer bidders.} each with valuation $v_i$ per click~\cite{Varian07}. Every slot $ 1 \leq j \leq s$ is associated with a click-through rate (CTR) $x_j > 0$, where $x_j \geq x_{j+1}$. For mathematical convenience, we define $x_j=0$ for every $j > s$. 
Throughout the paper we make the simplifying assumptions that CTRs are strictly decreasing (i.e., $x_j > x_{j+1}$), and that $v_i\neq v_j$ for all $i\neq j$. We denote by bold letter the corresponding vectors of valuations, CTRs, and bids (e.g. $\vec b=(b_1,\ldots,b_n)$).

A bidder $i$ that has been allocated slot $j$ gains $v_i$ per click (regardless of the slot), and is charged $p_j$ per click.
Thus, her total utility is given by $u_i = (v_i-p_j)x_j$.

\subpar{VCG}
In the VCG mechanism every bidder $i$ submits a bid $b_i$, and the mechanism allocates the $j$'th slot, $j=1, \ldots, s$, to the $j$'th highest bidder.
Each bidder $j$ is charged (per click) for the ``harm'' she poses to the other bidders, i.e., the difference between the welfare of bidders $k \neq j$ if $j$ is omitted and their welfare when $j$ exists.

It is well known that the VCG mechanism is \emph{truthful}, meaning that reporting true valuations $b_j=v_j$ is a (weakly) dominant strategy for all bidders. In particular, it is a Nash equilibrium.

Suppose that bidders' valuations are sorted in non-increasing order.
Assuming truthful bidding (i.e. $b_j=v_j$ for all $j$), each bidder~$i\leq s$ is allocated slot $i$, and pays 

\labeq{VCG_pay}
{p^{VCG}_i = \sum_{s+1 \geq j \geq i+1} \frac{x_{j-1}-x_j}{x_i} \cdot v_j.}

\subpar{GSP}
In the GSP auction, slot $j$ is given to the $j$'th highest bidder (as in the VCG auction).
Denote by $j$ the bidder who is getting slot $j$. 
The charge of bidder $j = 1, \ldots, s$ equals to the bid of the next bidder; i.e., $p_j = b_{j+1}$. For mathematical convenience, we define $b_{j+1}=0$  for $j \geq n$.

\subpar{GSP equilibria}
Varian~\cite{Varian07} identifies a set of natural Nash equilibria of the GSP auction, termed \emph{envy free NE} or \emph{Symmetric NE} (SNE), which are characterized by a set of recursive inequalities.
Varian shows that all SNE's satisfy some very convenient properties. First, in SNE no bidder wants to swap slots with any other bidder.\footnote{When swapping with a bidder in a worse slot, this requirement coincides with the one implied by NE. However when swapping with a bidder in a better slot, envy-freeness is slightly stronger.} Second, SNEs are efficient in the sense that bidders with higher valuations always bid higher (and thus get better slots). This allows us to assume that valuations are also sorted in non-decreasing order $v_1\geq v_2\geq\cdots\geq v_n$. Lastly, SNEs can be easily computed by a recursive formula, which makes them especially attractive for computerized and online settings.

The two equilibria that reside on the boundaries of the SNE set, referred to as \emph{Lower Equilibrium} (LE) and \emph{Upper Equilibrium} (UE), are of particular interest.
We denote the LE and UE profiles by $\vec b^L = (b_i^L)_{i \in N}$ and $\vec b^U=(b_i^U)_{i \in N}$, respectively.
The bids in the LE, for every $2 \leq i \leq s+1$, are given by
\begin{align*}
b^L_i x_{i-1}  =  v_i(\!x_{i-1}-x_i\!) + b^L_{i+1}x_i
 =  \!\!\!\!\!\sum_{s+1 \geq j\geq i} \!\!\!\!\!v_j(x_{j-1}-x_j).
\end{align*}
In particular, since CTRs are strictly decreasing, we get that $b_i > b_{i+1}$ for all $i\leq s$.
A central result by Varian~\cite{Varian07} is that the LE equilibrium induces payments, utilities, and revenue equal to those of the truthful outcome in VCG. It is therefore of great interest to compare the stability of these seemingly identical outcomes in both mechanisms.
%


The bids in the UE, for every $2 \leq i \leq s+1$, are given by
\begin{align*}
b^U_i x_{i-1}  =  v_{i-1}(\!x_{i-1}\!-\!x_i\!)\! +\! b^U_{i+1}x_i
 =  \!\!\!\!\!\!\! \sum_{s+1 \geq j\geq i}\!\!\!\!\!\!v_{j-1}(\!x_{j-1}\!-\!x_j\!).
\end{align*}

In the remaining of this section we measure the stability of the VCG and GSP mechanisms.
Our results indicate that while the mechanisms have seemingly identical outcomes, for many natural valuation and CTR functions, GSP is far more stable than VCG.

\subsection{Deviations in VCG}

Recall that the payment for bidder $i$ is a weighted average of reported (and by truthfulness, the actual) values of bidders $i+1 \leq j \leq s+1$ (see Eq.~\eqref{eq:VCG_pay}).


We next characterize the structure of a set of deviators $R$ of size $r$.
We say that a coalition $R$ of $r$ bidders has a \emph{potential to deviate} under VCG (or that it is a \emph{potential coalition}), if either: (a) the group $R$ contains exactly $r$  \emph{winners} (i.e., bidders that are allocated a slot $j \leq s$); or (b)  the set $R$ is composed of $t<r$ winners, the first loser, and the $r-t-1$ bidders that directly follow (i.e., bidders $s+1$ through $s+r-t$).

We denote the number of potential coalitions of size $r$ by $M_r$. We argue that it only makes sense to count potential coalitions when considering a deviation. 

To see why, note first that all bidders ranked $s+r$ or worse have no effect on the payment of any other bidder, and can be ignored.
Second, the bidders ranked $s+2,\ldots,s+r-1$ are only effective if they allow the bidder allocated slot $s+1$ to lower her bid.
Thus non-potential coalitions must contain at least one bidder that has no contribution at all to the deviation, and can therefore be ignored. Note for example that while adding dummy bidders (with valuation $0$) increases the total number of coalitions, the number of potential coalitions remains unchanged.

It is easy to verify that there are $\binom{s}{r}$ coalitions of type (a), and $\sum_{t=1}^{r-1}\binom{s}{t}$ coalitions of type (b). Thus $M_r = \sum_{t=1}^r \binom{s}{t}$.
Interestingly, in VCG every potential coalition can actually deviate. 

\begin{proposition}
\label{th:VCG_all_deviate}
Under the truthful equilibrium of VCG, denoted by $T$, any potential coalition has a deviation, i.e., $D_r(VCG,T) = M_r$ for all $2\leq r \leq s$.
\end{proposition}
\begin{proof}
Let $R$ be some potential coalition, and $i^* \in \argmin_{i\in R} v_i$. We call $i^*$ the \emph{indifferent bidder}.
Suppose that every agent $i\in R$ reports $v'_i$ so that $v_i >v'_i > v_{i+1}$.
Clearly, this has no effect on slot allocation.
In coalitions that include only winners, all the agents except agent $i^*$ (which is indifferent) pay strictly less than their original payments, as the payment monotonically depends on the valuations of the other members of $R$. In potential coalitions other type, where $R$ includes $t$ winners and $r-t$ losers, all $t$ winners strictly gain.
\end{proof}

\subsection{Deviations in GSP}
Since LE is a Nash equilibrium, we have that $\mc D_1(GSP,LE) = \mc{SD}_1(GSP,LE) = 0$.
In fact, as in the VCG mechanism, no coalition has a strict deviation from the LE profile in GSP. This statement is not as trivial in the GSP mechanism, but it follows from Lemma~\ref{lemma:all_coalitions} toward the end of this section.
The same analysis holds for the UE in GSP.
We next turn to evaluate the resistance of GSP to (non-strict) deviations, focusing on the lower equilibrium. As in the previous section, we only count potential coalitions as all other coalitions necessarily contain redundant participants.

\newpar{Pair deviations: characterization}
We begin by characterizing all deviations by pairs of agents.

\subpar{Lower equilibrium}
It is easy to see that for every $i\leq s$, the pair of agents $(i,i+1)$ (called \emph{neighbors}) can always (weakly) gain as a coalition, by having agent $i+1$ lowering her bid to $b'_{i+1}$, so that $b_{i+1} > b'_{i+1} > b_{i+2}$.\footnote{The assumption that CTRs are strictly decreasing is required here, as otherwise bidder $i+1$ may not be able to lower her bid.}
In this case, agent $i+1$ is not affected, but agent $i$ gains the difference $x_i(b_{i+1}-b'_{i+1}) > 0$. It is also clear that bidders ranked $s+2$ or worse can never be part of a deviating pair. In terms of the stability score, this means that
\vspace{-2mm}
 $$~~~~~~~~~~~~~~s\leq \mc D_2(GSP,LE) \leq M_2 = \binom{s+1}{2}.\vspace{-0mm}$$

Consider the pair of agents $(k, j)$, where $k < j \leq s+1$. We want to derive a sufficient and necessary condition under which the pair $(k,j)$ has a deviation.
A simple observation is that given some Nash equilibrium, for an agent $i$ to strictly gain by being allocated a new slot $i'\neq i$, the bid $b_{i'+1}$ must strictly decrease, since otherwise this would also be a deviation for $i$ as a single agent (in contradiction to equilibrium).
Therefore, either (1) $k$ moves to a worse slot $k'=j-1$, and $b'_{j} < b_{j}$; or (2) $j$ moves to a better slot $j'=k$, $k$ is pushed down to $k'=k+1$, and $b'_{k} < b_k$.
However, if $j$ gains in case (2), then this means she is envy in bidder $k$. This is impossible, as we assumed $\vec b$ is an SNE.
Thus, the only deviation is where $k'=j-1; j'=j$. Further, this is a deviation only if $b_{j-1} >  b'_{k} > b'_j \geq b_{j+1}$.
Note that: (i) $b'_k$ can get any value in this range without affecting the utility of $k$ or $j$, (ii) the utility of $j$ remains the same, and (iii) the most profitable deviation for $k$ is one in which $b'_j = b_{j+1}$ (breaking the tie in favor of $j$).

The discussion above establishes a necessary condition for a pair deviation, and asserts that in every pair deviation of $k,j$ only agent~$k$ can strictly gain, where $k<j$.
We next complete the characterization by establishing a sufficient condition for pair deviation.

For the following results, we denote $a= x_{j-1}-x_j$ (for our fixed $j$), and $w_i = \frac{x_{i-1}-x_i}{x_j}$ for all $i\leq s+1$.
\begin{lemma}
\label{lemma:condition_LE} Suppose that the pair $k,j$ deviates from LE, by moving agent~$k$ to slot $k'=j-1$.
Let $u(k),u'(k)$ be the utility of agent~$k$ before and after the deviation, then 
$$u(k) -u'(k)\geq\sum_{t=k+1}^{j-1} (x_{t-1} - x_t) (v_k-v_t) - a\cdot v_j + a  \sum_{i=j+1}^{s+1} w_i v_i.$$
Moreover, in the optimal deviation for agent~$k$ the last inequality holds with an equality.
\end{lemma}
\begin{proof} 
Suppose agent~$j$ lowers her bid to $b'_j = b_{j+1}+\eps$ where $\eps\geq 0$ (so $j$ keeps her slot).
For any $\vec x, \vec v$ the utility of agent~$k$ changes as follows:
\begin{align*}
&u(k) -u'(k)= (v_k-b_{k+1})x_k - (v_k-(b_{j+1}+\eps))x_{j-1} \notag \\
&= (x_k - x_{j-1}) v_k - \!\!\!\sum_{t=k+1}^{s+1}\!\! (x_{t-1} - x_t) v_t \\
&~~~~~~~~~~~~+ \! \sum_{i=j+1}^{s+1}\!\! \frac{x_{j-1}(x_{i-1} - x_i)}{x_j} v_i +\eps x_{j-1} \notag\\
&= \sum_{l=k+1}^{j-1} (x_{l-1} - x_l) v_k   - \sum_{t=k+1}^{j} (x_{t-1} - x_t) v_t  \\
&~~~~~~~~~~~~ + \(\frac{x_{j-1}}{x_j}-1\)  \sum_{i=j+1}^{s+1} (x_{i-1} - x_i) v_i  +\eps x_{j-1}\notag\\
&=\sum_{t=k+1}^{j-1} (x_{t-1} - x_t) (v_k-v_t) - (x_{j-1} - x_j) v_j \\
&~~~~~~~~~~~~+ \frac{x_{j-1}-x_j}{x_j}  \sum_{i=j+1}^{s+1} (x_{i-1} - x_i) v_i  +\eps x_{j-1}\notag \\
&= \sum_{t=k+1}^{j-1} (x_{t-1} - x_t) (v_k-v_t) \\
&~~~~~~~~~~~~- a\cdot v_j + a\!\!  \sum_{i=j+1}^{s+1} w_i v_i +\eps x_{j-1}. \notag 
\end{align*} 
The inequality follows since $\eps \geq 0$. In the optimal deviation $\eps=0$ in which case we get an equality.
Note that $\sum_{i=j+1}^{s+1} w_i v_i$ is a weighted average of valuations.  In particular, it is always between $v_{s+1}$ and $v_{j+1}$.
\end{proof}

As a direct corollary from Lemma~\ref{lemma:condition_LE}, we get that in LE the pair $k,j$ (where $k<j-1$), has a deviation \emph{if and only if}
\labeq{c_pair}
{\sum_{t=k+1}^{j-1} (x_{t-1} - x_t) (v_k-v_t) < a\cdot v_j - a  \sum_{i=j+1}^{s+1} w_i v_i.}

\subpar{Upper equilibrium}
It is easy to check that a similar characterization to Eq.~\eqref{eq:c_pair} applies to the UE. However, the conditions differ with respect to bidders that are two positions apart.
\begin{proposition}
\label{th:UE}
Given a UE, the pair of agents $i, i+2$ has a deviation for every $i<s$.
\end{proposition}
This result holds under all valuation and CTR functions; hence $\mc D_2(GSP,UE) \geq 2s-1$. This means that the UE may be slightly less stable than LE (whose stability is expressed in Theorem~\ref{th:convex}).
Yet, it is not too difficult to show that the number of pair deviations from UE and LE are asymptotically the same.
Therefore, in the remainder of this section we focus on stability scores of LE.

\newpar{Pair deviations: quantification}
It turns out that the asymptotic number of pair deviations strongly depends on the shape of both the CTR function and the valuation function. 
In particular, convexity (as well as concavity and $\beta$-convexity) will play a major role in our results.
Let $g_1,\ldots,g_m$ be a monotonically \emph{nonincreasing} vector.

 Similarly to the way defined convex cost functions in Section~\ref{SEC:RSG}, we say that $g$ is \emph{convex} if it has a decreasing marginal loss; i.e., $g_i - g_{i+1} \geq g_j-g_{j+1}$ for every $i<j$. Similarly, if $g$ has an \emph{increasing} marginal loss then it is \emph{concave}.

 Note that linear functions are both convex and concave. A special case of convexity (resp., concavity) is when the marginal loss decreases (resp., increases) exponentially fast.

Let $\beta>1$.
We say that $g$ is \emph{$\beta$-convex} if $g_{i-1} - g_{i} \geq \beta (g_i-g_{i+1})$ for every $i$.
Similarly, $g$ is said to be \emph{$\beta$-concave} if $\beta(g_{i-1} - g_{i}) \leq g_i-g_{i+1}$ for every $i$.
\footnote{Lucier et al.~\cite{LPT11} studied GSP auctions with \emph{well-separated} CTR functions, which is a closely related term. In particular, a $\frac{1}{\beta}$-well separated function is also $\beta$-convex.}

Intuitively, as either valuations or CTRs are ``more'' convex,\footnote{When referring to convexity of CTR/valuation functions, we only consider the first $s+1$ values.} a bidder who deviates by moving to a lower (i.e., worse) slot faces a more significant drop in her utility.
Thus we can hope that pairs that are sufficiently distant from one another will not be able to deviate jointly.
This intuition is further formalized and quantified in the remainder of this section.
For convenience, the results are summarized in Table~\ref{tab:m_results}.
\begin{table}
\begin{center}
\small{
\begin{tabular}{|l l|c|c|c|c|}
\hline
 \multicolumn{2}{|c|}{\multirow{2}{*}{\backslashbox{Valuations}{CTR}}}    & \multicolumn{2}{|c}{$\leftarrow$ concave $\rightarrow$} & \multicolumn{2}{c|}{$\leftarrow$ convex $\rightarrow$} \\

 & & $\beta$-concave & \multicolumn{2}{|c|}{Linear} & $\beta$-convex \!\!\\
  \hline
  \multicolumn{1}{|l}{concave} & \!\! $2$-concave \!\! &\rule{0pt}{2.5ex}
 All $\binom{s+1}{2}$ &  \multicolumn{2}{|c|}{\!\! All $\binom{s+1}{2}$ \!\!} & -\!\! \\
  \cline{2-6}
     \multicolumn{1}{|l}{} & \!\! Linear \!\! &  \rule{0pt}{2.5ex}
 $\Omega(s^2)$ &  \multicolumn{2}{|c|}{$\Theta(s\sqrt s)$ \!\!} & $ \!\! O(s\!\cdot\!\log_\beta s)$ \!\!\!\!\\
    \cline{2-6}
    \multicolumn{1}{|l}{convex} & \!\! $2$-convex \!\! &  - &  \multicolumn{2}{|c|}{$s$} & $s$ \!\!\\
    \hline
\end{tabular}
}
\end{center}
\caption{\label{tab:m_results}The table summarizes the number of pairs that have a deviation, i.e., $\mc D_2(GSP,LE)$. When one function is strictly concave and the other is strictly convex, the score may depend on the exact structure of both functions.}
\end{table}

The next proposition demonstrates that convexity induces greater stability.
\begin{theorem}
\label{th:convex} Suppose that both CTR and valuation functions are \emph{convex}. The number of pairs with deviations in the Lower equilibrium can be upper bounded as follows.
\begin{enumerate}[(A)]
	\item $\mc D_2(GSP,LE) = O(s\sqrt s)$.
	\item if CTRs are $\beta$-convex then $\mc D_2(GSP,LE)=O(s\log_\beta s)$.
	\item if valuations are $\beta$-convex, for any $\beta\geq 2$, then only neighbor pairs can deviate. I.e., $\mc D_2(GSP,LE) = s$.
\end{enumerate}
\end{theorem}
We present the proof of the first statement, so as to demonstrate the proof technique. 

\begin{proof}[of \ref{th:convex}(a)]
Recall that $a = x_{j-1}-x_j > 0$. A crucial observation is that $\sum_{i=k+1}^{s+1} w_i v_i$ is in fact a weighted average of valuations, where the weight $w_i$ is proportional to the difference $x_{i-1}-x_i$. Therefore this average is biased toward low values when CTR is convex, and toward high values when it is concave.

Also, since CTRs are convex, we have that for all $i<j$, $x_{i-1}-x_i \geq a$. Thus by Lemma~\ref{lemma:condition_LE},
\begin{align}
u(k) -& u'(k) \geq a\sum_{t=k+1}^{j-1} (v_k-v_t)    - a \cdot v_j  + a \sum_{i=j+1}^{s+1} w_i v_i \notag \\
&= a\(\sum_{t=k+1}^{j-1} (v_k-v_t)  +  \sum_{i=j+1}^{s+1} w_i v_i - v_j\) \notag \\
&\geq a\(\sum_{t=k+1}^{j-1} (v_k-v_t)  + \avg_{s+1\geq i \geq j+1}\!\!\!\!(v_i) - v_j\). \label{eq:avg}
\end{align}
Therefore, in order to prove that the pair $j,k$ can deviate, it is necessary to show
\vspace{-5mm}
\labeq{pair}
{~~~~~ \sum_{t=k+1}^{j-1}(v_k-v_t) < v_j - \avg_{s+1\geq i\geq j+1}v_{i}.}
We note that under linear CTRs, all inequalities become equalities (in which case Equation~\eqref{eq:pair} is also a sufficient condition).
Observe that closer pairs are more likely to deviate. E.g. for pairs s.t. $j=k+2$, it is sufficient that
$\displaystyle v_k-v_{k+1}  < v_{k+2} - \avg_{s\geq t'\geq k+3}v_{t'}$
to have a deviation.
Let $h=j-1-k\geq 1$, and $z = v_k-v_{j-1} = v_k - v_{k+h}$.

From convexity of $\vec v$ it holds that for all $h'<h$, $\frac{v_k-v_{k+h'}}{h'} \geq \frac{v_k-v_{k+h}}{h}=\frac{z}{h}$, thus for the LHS of Eq.~\eqref{eq:pair},

\labeq{v_avg}
{\sum_{t=k+1}^{j-1}(v_k-v_t) \geq \sum_{t=k+1}^{j-1}z\frac{t-k}{h} = \frac{z}{h}\frac{h(h+1)}{2} = \frac{h+1}{2}z.}

Bounding the RHS of Eq.~\eqref{eq:pair}, we have

\begin{align}
v_j-&\!\! \avg_{s+1\geq i\geq j+1} \!\! v_i \leq v_j -v_{\avg\{s+1\geq i\geq j+1\}}\tag{convexity of $\vec v$}\\
& \leq v_j - v_{\ceil{\frac{\rule{0pt}{1ex}j}{\rule{0pt}{1ex}2}+\frac{\rule{0pt}{1ex}s}{\rule{0pt}{1ex}2}}} = v_j - v_{\ceil{j+\frac{s-j}{\rule{0pt}{1ex}2}}} \notag\\
&\leq \!\!\!\!\sum_{i'=1}^{\ceil{(s-j)/2h}}\!\!\!\!\!(v_{j+(i'-1)h} - v_{j+i'h}) \leq \!\!\!\!\sum_{i'=1}^{\ceil{(s-j)/2h}}\!\!\!\!\!(v_k - v_{k+h}), 
\label{eq:RHS}
\end{align}
which is at most $\ceil{\frac{s-j}{2h}}z$.
By using the bounds we showed on both sides of the equation, condition~\eqref{eq:pair} implies $h+1 < \ceil{\frac{s-j}{h}}$,
which must be false whenever $h+1 = j-k > \sqrt s$. Therefore each winner $k\leq s$ can deviate with at most $\sqrt s$ other bidders, and there can be at most $s\sqrt s$ such pairs.
\end{proof}

It is evident from Theorem~\ref{th:convex}, that convexity can guarantee some level of stability, and further, that ``more'' convexity can induce more stability. Our next result complements this observation, by showing that \emph{concavity} of valuation and CTR functions affects stability in the opposite direction.

\begin{theorem}
\label{th:concave}
Suppose that both CTR and valuation functions are \emph{concave}. The number of pairs with deviations in the Lower equilibrium can be lower bounded as follows.
\begin{enumerate}[(A)]
	\item $\mc D_2(GSP,LE) = \Omega(s \sqrt s)$.
	\item if CTRs are $\beta$-concave for any $\beta>1$, then $\mc D_2(GSP,LE)=\Omega(s^2)$ (i.e. a constant fraction of all pairs).
	\item if valuations are $\beta$-concave, for any $\beta \geq 2$, then \emph{all} pairs can deviate.
I.e., $\mc D_2(GSP,LE) = \binom{s+1}{2} = M_2$.
\end{enumerate}
\end{theorem}

A linear function is both convex and concave. Therefore, in the special case where both CTRs and valuations are linear, we obtain an asymptotically tight estimation of $\mc D_2(GSP,LE)$.

\newpar{Deviations of more than two agents}
\label{SEC:more_than_pairs}
We first characterize the structure of such deviations.

\begin{lemma} \label{lemma:all_coalitions}
Suppose that $R\subseteq N$ is a coalition that gains by a deviation, and let $b_j,b'_j$ denote the bids of $j\in R$ before and after the deviation, respectively. Then the following hold:
\begin{enumerate}[(a)]
	\item There is at least one bidder $i^*\in R$ that does not gain anything from the deviation; this bidder is called the \emph{indifferent bidder}.
	\item There is at least one bidder $f\in R$ s.t. $R\setminus\{f\}$ still has a deviation; this bidder is called a \emph{free rider}. 
	\item For all $j\in R$, either $b'_j < b_j$, or the utilities of all agents in $R$ (including $j$) are unaffected by the bid of $j$.
\end{enumerate}
\end{lemma}

In order to prove the Lemma, we must show that the bidder that is ranked last among the deviators is an indifferent bidder ($i^*$). The free rider ($f$) is either the bidder that is ranked first among the deviators, or some bidder that is isolated of all other deviators.
In addition, it is shown that bidders that move to a better slot either strictly lose, or cause some other deviator to strictly lose.

\medskip

As a direct corollary of Lemma~\ref{lemma:all_coalitions}, given any coalition $R$ of size $\geq 3$, the coalition $R \setminus \{f\}$ can also deviate.
By induction, therefore, a coalition $R$ that can deviate always contains a pair that can deviate.
Moreover, by part (c) of Lemma~\ref{lemma:all_coalitions}, it follows that given a deviating coalition of size $\geq 2$, it can be extended by adding a bidder who does not change her bid.
As a result, a set $R$ can deviate if and only if it contains a pair that can deviate.
This crucial observation facilitates the computation of the number of deviations by coalitions of size $r$ for any $r \geq 3$.

Recall that $M_r$ denotes the number of potential coalitions of size $r$, and that under VCG auction all of these coalitions actually have a deviation. Clearly, $\mc D_r(GSP,LE)\leq M_r$. We next show how the accurate number of coalitions asymptotically depends on the size of the coalition $r$ and on the number of slots $s$.

\begin{proposition}
\label{th:LE_subsets}
If both CTRs and valuations are convex, then
\vspace{-2mm}
$$\mc D_r(GSP,LE) \leq M_r \cdot O\(\frac{r^2}{\sqrt s}\).$$
In contrast, if both CTRs and valuations are concave, then
$$\mc D_r(GSP,LE) \geq M_r \cdot d\cdot \(\!1-\exp\(\!-\Omega\(\frac{r\sqrt r}{\sqrt s}\)\)\)$$ for any positive constant $d<1$.
\end{proposition}
That is, at least in the convex case the number of potential deviations under GSP is significantly smaller than under VCG.

This result also establishes an almost sharp threshold for the case of linear CTRs and valuations.
In particular, for every $r \gg ~^{3}\!\!\!\sqrt s$, almost all coalitions of size $r$ can deviate, while the proportion of coalitions of size  $r  \ll ~^{4}\!\!\!\sqrt s$ that can deviate goes to 0 (when $r$ is fixed and as $s$ grows).

Proposition~\ref{th:LE_subsets} confirms that the GSP auction is far more stable than the VCG auction against collusions of relatively small coalitions (at least when CTR and valuations are convex). 

\section{Eliminating Group Deviations}
\label{SEC:RESERVE}
\subsection{VCG with a reserve price}
\label{SEC:reserve_VCG}

Consider a variant of the VCG mechanism that adds a fixed reserve price $c$.
That is, only bidders that reports a value of $c$ or higher get a slot, and payments are computed ignoring the other bidders (i.e. replacing their values with $c$). 
It is easy to verify that truth-telling remains a dominant strategy, and that Proposition~\ref{th:VCG_all_deviate} remains valid if the values of all bidders are strictly above $c$.
However, a bidder whose value is exactly $c$ will not join any coalition: by lowering her reported value she will lose her current slot for sure, whereas previously she enjoyed a positive utility.

Now, consider a VCG mechanism that chooses a reserve price as follows. With probability $q$, the reserve price is chosen randomly from a sufficiently large interval, 
and with probability $1-q$, it is set to 0.
Crucially, the probability distribution of the reserve price is common knowledge, but agents submit their reports before its realization is revealed. Let us denote the proposed mechanism by VCG$^*$. While the proposed adjustment seems small, it results in a dramatic increase of stability.
\begin{theorem}
\label{th:VCG_group}
If $s\geq n$, then truth-telling is a SSE in VCG$^*$.
\end{theorem}
\begin{proof}
First observe that VCG$^*$ is a lottery over strategyproof mechanisms, thus no single agent has an incentive to deviate unilaterally.
Suppose by way of contradiction that there exists a coalition that gains by a deviation, and let $R$ be such a coalition of minimal size.
Since $R$ is minimal, the indifferent agent $i^* \in  R$ (as defined in Prop.~\ref{th:VCG_all_deviate}) must lower her reported value, otherwise the coalition $R \setminus \{i^*\}$ can also deviate.
Assume, therefore, that $v'_{ i^*} = v_{ i^*} - \eps$ for some $\eps > 0$.
It is easy to verify that there is no outcome of the mechanism under which $i^*$ gains.
In contrast, there is a non-zero probability that $c$ is chosen in the range $(v'_{i^*},v_{i^*})$, in which case the utility of $i^*$ becomes 0, compared to $(v{i^*} - c)x_{i^*} > 0$ under truth-telling.
Therefore, agent $i^*$ loses in expectation, contradicting the existence of a coalition $R$.
The assertion of the theorem is established.
\end{proof}
By the last theorem, VCG$^*$ guarantees stability whenever $n \leq s$.\footnote{The proof in fact shows a stronger result: truth-telling is a SSE in \emph{dominant strategies}. Thus VCG$^*$ is \emph{group-strategyproof}.}
However, if $s < n$ the bidder ranked $s+1$ can serve as the indifferent bidder of any coalition. Consequently, VCG$^*$ does not posses a SSE.
That is, since the utility of agent $s+1$ is always 0, she will not be discouraged by the random reserve price, even when her reported value falls below the reserve price.

In order to deal with the lack of slots (i.e., the case in which $s \leq n$), we introduce a modified VCG$^*$ mechanism, which always induces truth-telling as a SSE.

Consider the following modification to VCG$^*$, termed VCG$_{\lambda}^{*}$.
Let $0 < \lambda < \frac{1}{n}$. Given some slot $j\leq s$ with a CTR of $x_j>0$, it is allocated to the bidder that is ranked $j$ with probability $1-\lambda$, and is allocated to the bidder that is ranked $s+1$ with probability $\lambda$.
This modification effectively creates a new slot $s+1$, whose expected CTR is $\lambda x_j$, whereas the new (expected) CTR of slot $j$ becomes $(1-\lambda) x_j$.
This procedure can be applied to the desired additional $n-s$ slots.
In particular, a possible instantiation is where the new expected CTR of position $s$ will be $(1-(n-s)\lambda)x_s$, and there will be $n-s$ new slots with an expected CTR of $\lambda x_s$.
Since the new auction has $n$ slots, the mechanism VCG$^*$ can be performed to eliminate all coalitional deviations.

The careful reader will notice that by changing the CTRs, the equilibrium in the new auction may change.
However, as long as the order of the slots is preserved, the equilibrium allocation is not affected, and this is ensured by satisfying $\lambda < \frac{1}{n}$. Moreover, the new payment differs from the original payment by at most $v_1\cdot n \cdot \lambda$; thus for a sufficiently small $\lambda$ the difference is negligible.
As a result, we get the following corollary.

\begin{corollary}
\label{th:VCG_group_all}
Truth-telling is a SSE in mechanism VCG$_{\lambda}^{*}$ for every $0 \!<\! \lambda \!<\! \frac{1}{n}$.
Moreover, the payments and revenue of VCG$_{\lambda}^{*}$ can be arbitrarily close to the payments and revenue of VCG.
\end{corollary}

\subsection{GSP with a reserve price}
\label{SEC:reserve_GSP}
As evident from the results in the last section, stability of the VCG mechanism is significantly increased by augmenting the mechanism with a random reserve price and additional subtle randomization.
It might be tempting to apply the same technique to the GSP mechanism, in an attempt to increase its stability, while maintaining the possibility to achieve a higher revenue than VCG.
Unfortunately, this approach fails since (in contrast to VCG) adding a reserve price does not preserve its original set of equilibria.

To see this, consider a GSP mechanism with a fixed reserve price $c$.
Bidder $i$ is affected by the reserve price if either: (I)  $v_i > c > b_i$, in which case bidder $i$ has an incentive to raise her bid, as otherwise she will lose the slot; or (II) $v_i < c < b_i$, in which case she has an incentive to lower her bid, as otherwise she will pay more than the slot's worth to her.
In both cases it follows that the modified GSP mechanism no longer preserves the SNE properties characterized by Varian (even with respect to unilateral deviations).
The reason for the difference between VCG and GSP is that VCG induces truthful revelation in equilibrium; hence cases (I) and (II) suggested above cannot be realized.
%
%
%

\section{Discussion and future work}
\label{SEC:discussion}
Our main contribution in this paper is the introduction of \emph{stability scores} --- a new stability measure for game equilibria.
We demonstrated how stability scores can be used to compare equilibria in congestion games and to draw qualitative results regarding properties of the game and the profiles that increase coalitional stability.


\subpar{Auctions} Our results indicate that for a prominent class of CTR and valuation functions, GSP is far more stable than VCG.\footnote{Empirical studies indicate that CTRs on common platforms are indeed convex, see \cite{Atlas_report}.}

It is known that the LE of GSP generates exactly the same revenue as VCG, and any other SNE of GSP generates an even higher revenue.
This may suggest that GSP is better than VCG with respect to both revenue and stability.
However, a relatively simple modification to the VCG mechanism 
induces a randomized mechanism that eliminates all coalitional deviations, thus turning it into a highly stable mechanism. 
 An open question is whether our results still hold when ads' quality is also considered (see~\cite{Varian07}).

\subpar{Equilibria selection and mechanism design} 
Analysis of stability scores can be applied to various games and mechanisms.
In particular, in games that have multiple Nash equilibria such analysis can aid in selecting the an equilibrium. Understanding how coalitional stability is affected by properties of the game will help us to play better as players, and to create better games as designers.

\subpar{Toward a realistic picture of coalitional stability}
 Solution concepts such as $\eps$-NE (or $\eps$-SE) quantify the benefit an agent or coalition can get from a deviation. Therefore they offer stability under a relaxed notion of self-interest (i.e. agents will only bother to deviate for some substantial gain). In contrast, stability scores still assume purely self-interested agents, but relax a different aspect of coalitional rationality.  Practical limitations on information, communication or trust may mean that a coalition of agents will not collude even if they have a potentially high incentive to do so. 
Other models such as Myerson's~\cite{Myerson77} assume that limitations on collusion are given in an explicit and structured form.
 
 In future research we may wish take a combined approach to coalitional stability, considering both known and unknown limitations on collusion, possibly attributing more importance to coalitions with a stronger incentive to deviate.
Such models will enable us to better predict realistic outcomes of games, and to improve the mechanisms we design.




\bibliography{stability_scores}
\bibliographystyle{plain}

\onecolumn
\appendix
\include{appendix}

\end{document}

%% file: appendix.tex
\newcommand{\apxsection}[1]{%
  \section[#1]{Appendix: \centering #1}}
  \def\proofholder{Proof }
\apxsection{Proofs of {Section~\ref{SEC:RSG}}}
\label{apx:RSG}

\begin{rprop}{th:cong_two_hat_a} $\mc{D}_2(G,\hat{\vec a})=\Theta\(\frac{qn^2}{m^2}\)$. 
\end{rprop}
 \begin{proof}
  The first observation is that any NE profile must be almost-balanced, in the sense that every resource has $\floor{n/m}$ agents (vacant) or $\ceil{n/m}$ agents (full). Note that there are exactly $q$ full resources in each iteration. 
  
  The second observation is that a pair has a strict deviation if and only if they share a full resource in both iterations. Then one agent can switch to a vacant resource in the first iteration, and the other can do the same in the second iteration. In each iteration one of them strictly gains and the other is unharmed.
  
It follows that in $\hat{\vec a}$ agents play the same partition in both iterations, and every pair that is in a full resource can deviate. Since there are $q$ full resources, there are $q{ \ceil{n/m} \choose 2} =\Theta\(\frac{qn^2}{m^2}\)$.
\end{proof}

\begin{rprop}{th:cong_two_a_star}
\begin{enumerate}[(a)]
	\item $\mc{D}_2(G,\vec a^*)=O\( \frac{n^2}{m^2}\)$.
	\item if $n<m^2$ then $\mc{D}_2(G,\vec a^*)=0$, i.e. $\vec a^*$ is $2$-SE.
	\item if $q\leq m/2$, then  $\mc{D}_2(G,\vec a^*)=0$.
\end{enumerate}
\end{rprop}
\begin{proof}
  If $n<m^2$, then we show that $\vec a^*$ is a $2$-SE profile. Let $\vec A=\(A_1,A_2,\ldots,A_m\)$ be any almost-balanced partition in the first iteration. That is, $A_i$ contains the ($\ceil{n/m}$ or $\floor{n/m}$) agents that select resource $i$ in the first iteration. Assume each $A_i$ is ordered as a vector (arbitrarily). Let $\overline A$ be vector of size $n$, created by concatenating the vectors $A_1,\ldots, A_m$. We construct the partition in the second iteration $\vec B$, by adding each agent $\overline A(j)$ to resource ($j \mod m$). Since every $|A_i|\leq m$, all agents in $A_i$ end up in different resources in the second iteration. Thus $\mc{D}_2(G,\vec a^*)=0$.
 
 If $n > m^2$ and $q > m/2$, then there is at least one resource with $\geq m+1$ agents. By pigeon hole, at least two of these agents share a resource in the second iteration, thus $\mc{D}_2(G,\vec a^*)\geq 1$. However we can still upper bound the stability score of $\vec a^*$. Indeed, take any vector $A_i$, and divide it to subvectors $A_{i1},A_{i2},\ldots$, each of size $m$. We now create the partition $\vec B$ as described in the previous paragraph. As $|A_i|$ may be more than $m$, it is possible that two agents from $A_i$ now share a resource in $\vec B$. However if two agents belong to the same \emph{subvector} $A_{i,t}$, they must be in distinct resources in $\vec B$, and thus cannot deviate. Also, every $j\in A_{i,t}$ shares a resource in $\vec B$ with at most 1 other agent from each other subvector $A_{i,t'}$. Thus the number of pairs in $A_i$ shat share a resource in $\vec B$ is at most ${\ceil{|A_i|/m} \choose 2}$ (for example, $B_1$ contains the first agent from each set $A_{1,t}$, one agent from each $A_{2,t}$, etc.). However, not all of these pair can deviate. It is necessary that the resource shared in the first step is full (i.e. $|A_i|=\ceil{n/m}$), and also the shared resource in the second step. Thus, only a fraction of $q/m$ of the pairs end up in a full resource in $\vec B$. Thus for every full resource $i$, we have at most $\ceil{|A_i|/m}$ agents sharing a resource in $\vec B$. Summing the pairs from $A_i$ over the $q$ full resources of $\vec B$, we have (at most)
 $$q {\ceil{|A_i|/m} \choose 2} = \Theta\(q \(\frac{n}{m^2}\)^2\)$$
 deviating pairs, and the total number of deviating pairs in all $q$ full resources of $\vec A$ is
 \begin{align*}
 \mc{D}_2(G,\vec a^*)\leq q\cdot \Theta\(q \(\frac{n}{m^2}\)^2\) = \Theta\(\frac{q^2 n^2}{m^4}\) = \Theta\(\frac{n^2}{m^2}\).
 \end{align*} 
 
 For the last case, suppose that $q<m/2$. We take one agent from each full resource in $\vec A$, and move it to a (distinct) empty resource to create $\vec B$. Thus there is no resource that is full in both iterations. Hence the only agents that belong to a full resource (and thus may have an opportunity to gain) in both iterations are the ones we moved. None of these agents shares a resource with any other agent twice, and therefore no pair deviation is possible.
\end{proof}

\begin{rprop}{th:cong_random} Let $G$ be an SRSG with $k$ steps, and $\vec a$ be a random NE in $G$. Denote $r=\frac{q(k-1)}{m^2}$,  then $\mc{D}_2(G,\vec a) \cong {n \choose 2}\(1-(1+r)e^{-r}\)$.
\end{rprop}
\begin{proof}
Let $(1,2)$ be a random pair of agents. In each iteration, they share a resource w.p. of $\frac{1}{m}$. Also, if they do share a resource, this resource is full w.p. of $\frac{q}{m}$, thus they have a probability of $\alpha=\frac{q}{m^2}$ to share a full resource. $(1,2)$ can deviate iff they share at least two full resources. Equivalently, they do not have one iff they share exactly 0 or 1 full resource, which occurs at probability of 
\begin{align*}
\beta &=  (1-\alpha)^k + k\cdot\alpha(1-\alpha)^{k-1} \\
& = (1-\alpha)^{k-1} (1-\alpha +k\alpha) \cong e^{-\alpha(k-1)} (1+\alpha(k-1)) \\
& =  e^{-r} (1+r). \tag{as $r=\frac{q(k-1)}{m^2}=\alpha(k-1)$}
\end{align*}
Since every pair \emph{does not} have a deviation w.p. $\beta$, the expected number of pair deviations is ${n \choose 2}(1-\beta) = {n \choose 2}\(1-(1+r)e^{-r}\)$.
\end{proof}

\apxsection{Proofs of Section~\ref{SEC:AUCTIONS}}
\label{apx:auctions}

\subsection{Characterizing pair deviations}
\label{apx:pair_char}

%
%

\begin{lemma}
\label{lemma:condition_UE}
The following condition is both necessary and sufficient for the pair $k<j-1$ to have a deviation in UE:
$$
\sum_{t=k+1}^{j-1} (x_{t-1} - x_t) (v_k-v_{t-1}) < a\( v_{j-1} -   \sum_{r=j+1}^{s+1} w_r v_{r-1}\),
$$
where according to our notations  $a=x_{j-1}-x_j$, and $w_r = \frac{x_{r-1}-x_r}{x_j}$.
\end{lemma}
\begin{proof}
The proof is very similar to that of Lemma~\ref{lemma:condition_LE}. Let $u(k),u'(k)$ be the utility of agent~$k$ before and after the deviation.
Recall that the best thing that the pair $k<j-1$ can do, is that $j$ reports $b'_j = b_{j+1}$, and $k$ reports $b'_k = b_j$ (i.e. takes slot $j-1$). The new utility of $k$ in this case is $u'(k) = (v_k-b_{j+1})x_{j-1}$.
For any $\vec x, \vec v$ the utility of $k$ changes as follows:
\begin{align*}
u(k) -&u'(k)= (v_k-b^U_{k+1})x_k - (v_k-b^U_{j+1})x_{j-1} \notag\\
= &(x_k - x_{j-1}) v_k - \sum_{t=k+1}^{s+1} (x_{t-1} - x_t) v_{t-1} +\frac{x_{j-1}}{x_j}  \sum_{r=j+1}^{s+1} (x_{r-1} - x_r) v_{r-1} \notag\\
=& \sum_{l=k+1}^{j-1} (x_{l-1} - x_l) v_k   - \sum_{t=k+1}^{j} (x_{t-1} - x_t) v_{t-1}  + (\frac{x_{j-1}}{x_j}-1)  \sum_{r=j+1}^{s+1} (x_{r-1} - x_r) v_{r-1} \\
=&\sum_{t=k+1}^{j-1} (x_{t-1} - x_t) (v_k-v_{t-1}) - (x_{j-1} - x_j) v_{j-1} + \frac{x_{j-1}-x_j}{x_j}  \sum_{r=j+1}^{s+1} (x_{r-1} - x_r) v_{r-1} \\
=& \sum_{t=k+2}^{j-1} (x_{t-1} - x_t) (v_k-v_{t-1})  - a \cdot v_{j-1} + a\sum_{r=j+1}^{s+1} w_r v_{r-1}
\end{align*}
\end{proof}

From Lemma~\ref{lemma:condition_UE} we can derive bounds on stability scores that are asymptotically equal to the ones we derived for LE. 

\begin{rprop}{th:UE}
Given a UE, the pair of agents $i, i+2$ has a deviation for every $i<s$.
\end{rprop}
\begin{proof}
We take Lemma~\ref{lemma:condition_UE}, and substitute $j$ with $k+2$. Then
\begin{align*}
u(k) -u'(k)&=\sum_{t=k+1}^{k+1} (x_{t-1} - x_t) (v_k-v_{t-1}) - a\cdot v_{k+1} + a  \sum_{r=k+2}^{s} w_{r+1} v_r \\
& = (x_{k} - x_{k+1}) (v_k-v_{k})  + a  \sum_{r=k+2}^{s} w_{r+1} v_r - a\cdot v_{k+1} \leq 0 + a\(v_{k+2} -v_{k+1} \) < 0. \tag{since $a>0$}
\end{align*}
Thus $u'(k) > u(k)$ and agent~$k$ strictly gains by deviating with $j=k+2$.
\end{proof}

\subsection{Counting pair deviations}
\label{apx:pair_count}
%


\begin{rtheorem}{th:convex} Suppose that both CTR and valuation functions are \emph{convex}. The number of pairs with deviations in the Lower equilibrium can be upper bounded as follows.
\begin{enumerate}[(A)]
	\item $\mc D_2(GSP,LE) = O(s\sqrt s)$.
	\item if CTRs are $\beta$-convex then $\mc D_2(GSP,LE)=O(s\log_\beta s)$.
	\item if valuations are $\beta$-convex for any $\beta\geq 2$, then only neighbor pairs can deviate. I.e., $\mc D_2(GSP,LE) = s$.
\end{enumerate}
\end{rtheorem}
Theorem~\ref{th:convex}(a) is proved in the main text.

\begin{proof}[of \ref{th:convex}(B)]
W.l.o.g. $x_s=1$.
As in the previous proof, we denote $a=x_{j-1}-x_j \geq \beta^{s-j}(x_{s-1}-x_s) = \beta^{s-j}(\beta-1)$. We can now rewrite differences between CTRs as $x_{i-1}-x_i \geq \beta^{j-i} a$ for all $i<j$. Continuing from Lemma~\ref{lemma:condition_LE},
\begin{align}
u(k)-u'(k)  &\geq\sum_{t=k+1}^{j-1} \beta^{j-t} a (v_k-v_t) - a\cdot  v_j + \frac{a}{x_j}  \sum_{r=j+1}^{s+1} (x_{r-1} - x_r) v_r \notag \\
&\geq a\(\sum_{t=k+1}^{j-1} \beta^{j-t} (v_k-v_t) -  v_j +  \avg_{s+1\geq r \geq j+1}\!\!\!\!v_r\), \label{eq:ex_avg}
\end{align}
where the inequality follows from the convexity of $\vec x$.
Thus we replace condition~\eqref{eq:pair} from the linear CTR case with
\labeq{ex_pair}
{\sum_{t=k+1}^{j-1} \beta^{j-t} (v_k-v_t) \geq  v_j -  \avg_{s+1\geq r \geq j+1}\!\!\!\!v_r,}
and make a similar analysis. Let $h,z$ as in the linear case, then
\begin{align}
\sum_{t=k+1}^{j-1} \beta^{j-t} (v_k-v_t) &\geq \sum_{t=k+1}^{j-1} \beta^{j-t} z \frac{t-k}{h} \tag{as in \eqref{eq:v_avg}}\\
&= \sum_{t=1}^{h} \beta^{h+1-t} z \frac{t}{h} = \frac{z}{h}\beta^{h+1}\sum_{t=1}^{h} \beta^{-t} t > \frac{z}{h}\beta^{h+1} \beta^{-1} = \frac{z}{h}\beta^{h}, \label{eq:exp}
\end{align}
Suppose now that $h > \log_\beta\frac{s-j}{2}$, then from Eq.~\eqref{eq:RHS}
$$\sum_{t=k+1}^{j-1} \beta^{j-t} (v_k-v_t) \geq \frac{z(s-j)}{2h} \geq  v_j -  \avg_{s+1\geq r \geq j+1}\!\!\!\!v_r,$$
which means no agent gains from the deviation.

Thus each bidder can find at most $O(\log_\beta s)$ other bidders to collaborate with, or $O(s \log_\beta s)$ pairs in total.

For tightness, assume that valuations are linear. In this case, all inequalities except \eqref{eq:exp} become equalities. Now take any pair such that $j< s/2;\ 2h< \log_\beta\frac{s-j}{2}$. Then we have
\begin{align*}
 h & < \log_\beta\frac{s-j}{2} - h < \log_\beta\frac{s-j}{2} - \log_\beta h = \log_\beta\frac{s-j}{2h}& \Rightarrow\\
\sum_{t=k+1}^{j-1} \beta^{j-t} (v_k-v_t) &= \sum_{t=1}^{h} \beta^{h+1-t} z \frac{t}{h} = \frac{z}{h}\beta^{h+1}\sum_{t=1}^{h} \beta^{-t} t \\
&< \frac{z}{h}\beta^{h+1} h\beta^{-1} = z \beta^{h} < z \frac{s-j}{2h} = v_j -  \avg_{s+1\geq r \geq j+1}\!\!\!\!v_r,
\end{align*}
and $k$ strictly gains by manipulating with $j$. Moreover, there are at least $\frac{s}{2}\frac{\log_\beta\frac{s}{4}}{2} =\Omega(s \log_\beta s)$ such pairs, thus our bound is tight.
\end{proof}

%

\begin{proof}[of \ref{th:convex}(C)]
Let any $k,j$ such that $j\geq k+2$.
\begin{align*}
\sum_{t=k+1}^{j-1}(v_k-v_t) &\geq v_k-v_{k+1} \geq 4( v_j - v_{j+1}) \geq 2 v_j \geq v_j - \avg_{s\geq t'\geq j+1}v_{t'}.
\end{align*}
Then by condition\eqref{eq:pair}, the pair $k,j$ cannot deviate.
\end{proof}

\begin{rtheorem}{th:concave}
Suppose that both CTR and valuation functions are \emph{concave}. The number of pairs with deviations in the Lower equilibrium can be lower bounded as follows.
\begin{enumerate}[(A)]
	\item $\mc D_2(GSP,LE) = \Omega(s \sqrt s)$.
	\item if CTRs are $\beta$-concave for any $\beta>1$, then $\mc D_2(GSP,LE)=\Omega(s^2)$.
	\item if valuations are $\beta$-concave, for any $\beta \geq 2$, then \emph{all} pairs can deviate.
I.e., $\mc D_2(GSP,LE) = {{s+1} \choose 2} = M_2$.
\end{enumerate}
\end{rtheorem}

\begin{proof}[\proofholder of \ref{th:concave}(a)]
Consider the proof of Theorem~\ref{th:convex}(A).
All the weak inequalities in the proof follow directly either from the convexity of $\vec x$, or from the convexity of $\vec r$. If both functions are concave, all weak inequalities are reversed (rounding expressions down rather than up).
Therefore, a pair $k,j=k+h+1$ can deviate whenever
$$h+1 < \floor{\frac{s-j}{h}}.$$
To see that there are $\Omega\(s\sqrt s\)$ such pairs, consider for example all pairs where $j<s/2; h<\sqrt{s/4}$.
\end{proof}

%
\begin{proof}[\proofholder of \ref{th:concave}(B)]
Consider Equation~\eqref{eq:ex_pair} in the proof of Theorem~\ref{th:convex}(B). As $\vec x$ is now concave, rather than convex, we have $x_{t-1}-x_t \leq \beta^{t-j}(x_{j-1}-x_j)$ for all $t<j$, and we should reverse the inequalities \eqref{eq:ex_avg} and \eqref{eq:ex_pair}. We get the following condition:
\labeq{ex_pair_concave}
{\sum_{t=k+1}^{j-1} \beta^{t-j} (v_k-v_t) <  v_j -  \avg_{s+1\geq r \geq j+1}\!\!\!\!v_r.}
Whenever condition~\eqref{eq:ex_pair_concave} holds, deviation of $k,j$ is guaranteed to succeed.
 Now, let $h=j-k-1$ as in previous sections. We show that each of the top $(1-\frac{1}{\beta}) \frac14  s$ bidders can deviate with any bidder above her (note that this means that there is a constant fraction of the total number of pairs that can deviate).
 We first upper bound the LHS:
 \begin{align*}
 \sum_{t=k+1}^{j-1} \beta^{t-j} (v_k-v_t) &\leq  (v_k-v_{j-1}) \sum_{t=k+1}^{j-1} \beta^{t-j}  = (v_k-v_{j-1}) \sum_{t=1}^{j-k-1} \beta^{t} \\
 &<  (v_k-v_{j-1}) \sum_{t=0}^{\infty} \beta^{-t} \leq  (v_k-v_{j-1}) \frac{1}{1-\frac{1}{\beta}} \leq a\cdot h \frac{1}{1-\frac{1}{\beta}} \\
 & \leq a \((1-\frac{1}{\beta})\frac14  s\) \frac{1}{1-\frac{1}{\beta}}  = \frac14 \cdot s \cdot a 
 \end{align*}
For the RHS, we have
\begin{align*}
  v_j -  \avg_{s+1\geq r \geq j+1}\!\!\!\!v_r &\geq v_j - v_{\frac{j+s+1}{2}} \geq v_j - v_{s/2}  \geq a \(\frac{s}{2} - j\) \geq a \(\frac{s}{2} - \frac14 \beta s\) \geq \frac14 \cdot s \cdot a. \tag{$\vec v$ is concave}
\end{align*}
We therefore have that for all $k<j< (1-\frac{1}{\beta})\frac14 s$, condition~\eqref{eq:ex_pair_concave} holds. Since $\beta>1$ then $(1-\frac{1}{\beta})>0$, and therefore there are $\Omega(s^2)$ such pairs, where the constant depends on $\beta$. For example, for $\beta=2$, there are  at least ${\floor{\frac18 s} \choose 2} > \frac{1}{100}s^2$ deviating pairs.
\end{proof}

\begin{proof}[\proofholder of \ref{th:concave}(C)]
By Equation~\eqref{eq:pair}, the pair $j,k$ can deviate if
\begin{equation*}
\sum_{t=k+1}^{j-1}(v_k-v_t) < v_j - \avg_{s+1\geq r\geq j+1}v_{r}.
\end{equation*}
Since $v_{i+1} - v_{i+2} > 2 (v_{i} - v_{i+1})$ for every $i$, for every $t > k$ it holds that
$$
v_k - v_t < \frac{v_{t} - v_{t+1}}{2^{t-k}}.
$$
We get
\begin{align*}
\sum_{t=k+1}^{j-1}&(v_k - v_t) =  (v_k - v_{k+1})(j-k-1) + (v_{k+1} - v_{k+2})(j-k-2) + \cdots + (v_{j-2} - v_{j-1})\\
 < & \frac{v_{j-2} - v_{j-1}}{2^{j-k-2}}(j-k-1) +
\frac{v_{j-2} - v_{j-1}}{2^{j-k-3}}(j-k-2)  +  \cdots + \frac{v_{j-2} - v_{j-1}}{2} + (v_{j-2} - v_{j-1})\\
 = & (v_{j-2} - v_{j-1}) \sum_{t=0}^{j-k-2} \frac{t+1}{2^t} <  (v_{j-2} - v_{j-1}) \(\sum_{t=0}^{j-k-2} \frac{t}{2^t} + \sum_{t=0}^{j-k-2} \frac{1}{2^t}\) <  (v_{j-2} - v_{j-1})(2+2)\\
 < & v_{j} - v_{j+1} <  v_{j} - \avg_{s+1\geq r\geq j+1}v_{r}.
\end{align*}
This establishes the statement of the proposition.
\end{proof}


\subsection{Counting deviations of large coalitions}
\label{apx:large}
\begin{rlemma}{lemma:all_coalitions}
Suppose that $R\subseteq N$ is a coalition that gains by a deviation, and let $b_j,b'_j$ denote the bids of $j\in R$ before and after the deviation. Then the following hold:
\begin{enumerate}
	\item There is at least one bidder $i^*\in R$ that does not gain anything from the deviation (an indifferent bidder). Moreover, the slot allocated to $i^*$ is not affected.
	\item There is at least one bidder $f \in R$ that does not contribute anything to the deviation (a ``free rider''). That is, the utility of all bidders in $R\setminus \{f\}$ does not decrease if $f$ bids her equilibrium bid, and at least one $j\in R\setminus\{f\}$ still gains.
	\item For all $j\in R$, either $b'_j < b_j$, or the utilities of all agents in $R$ (including $j$) are unaffected by the bid of $j$. 
\end{enumerate}
\end{rlemma}
\begin{proof} We prove each property separately.

\newpar{Indifferent bidder}
First consider the bidder $i^*\in R$ that is ranked last after the deviation, and let $i'$ be the new slot allocated to $i^*$.
Clearly $b_{i'+1}$ did not change, and thus if $i^*$ gains she would also gain by deviating unilaterally to $b'_{i^*} = b_{i'+1}+\eps$. Therefore $i^*$ is indifferent. Note that by our assumption that the game is generic, $i'=i^*$, or otherwise bidder $i^*$ would strictly lose.

\newpar{Lowering bids}
Suppose that $k\in R$ strictly gains by bidding $b'_k$ and moving to some slot $i$. Let $k^*$ be the bidder such that $b_{k^*}<b'_{k}$, and maximal in that condition (i.e. the bidder located directly below the new slot of $k$). Then either: (i) $k^*\in R$ and $b'_{k^*} < b_{k^*}$; or (ii) there is some bidder $t\in R$ such that $t<k$ (i.e. $b_t > b_k$), but after the deviation $b'_t < b'_k$. Let $t^*$ be the bidder $t$ with the lowest $b'_t$. If neither of (i),(ii) holds, then $k$ is allocated the same slot or worse, and pays at least as before.

Assume that $b'_k > b_k$.
If case (ii) holds, then $t^*$ strictly loses, or otherwise she would weakly gain by bidding $b'_{t^*}$ in a single deviation. Otherwise, note that $k$ itself does not gain, and consider some $j\in R$. Either $j<k$, $j$ remains above $k$, or $j>k$ and remains below $k$. In both cases $j$ is unaffected, unless $b'_j<b_k$ and maximal in that condition, in which case $j$ strictly loses by the move of $k$.

\newpar{Free rider} If $R$ contains a pair of neighbors, this pair has a deviation regardless of the actions of all other bidders, and we can clearly remove any bidder that is not a part of this pair. Assume therefore that $R$ do not contain a pair of neighbors.

Consider the bidder $f\in R$ that is ranked first among all bidders of $R$ (after the deviation), and denote her new slot by $f'$. Clearly $f$ does not contribute do any other bidder in $R$. $R\setminus \{f\}$ still has a deviation (i.e. do exactly what they did when $f$ was part of the coalition), unless $f$ is the only bidder that strictly gains by the deviation of $R$. Suppose we are in the latter case. According to our generic games assumption, bidders that do not gain must keep their slots, and by the previous paragraph, for all $k\in R\setminus \{f\}$, $b_k \geq b'_k\geq b_{k+1}$. Consider  $t\in R$, s.t. $t\neq f'+1$ (there must be such $t$, as $|R|\geq 3$. If $t=1$ bidder $t$ is a free rider and we are done, thus assume $t>1$.

Since $R$ contains no neighbors, the bidder in slot $t-1$ is not in $R$, and therefore the coalition $R\setminus \{t\}$ still has a deviation.
\end{proof}

\begin{rprop}{th:LE_subsets}
If both CTRs and valuations are convex, then
\vspace{-1mm}
$$\mc D_r(GSP,LE) \leq M_r \cdot O\(\frac{r^2}{\sqrt s}\).$$
In contrast, if both CTRs and valuations are concave, then
$$\mc D_r(GSP,LE) \geq M_r \cdot d\cdot \(\!1-\exp\(\!-\Omega\(\frac{r\sqrt r}{\sqrt s}\)\)\)$$ for any positive constant $d<1$.
\end{rprop}
%

\begin{proof} [\proofholder of Proposition~\ref{th:LE_subsets}, upper bound]
Recall that we only consider the top $s+r-1$ bidders.
The crucial observation is that a coalition $R$ can deviate iff it contains a pair that can deviate. This follows directly from Lemma~\ref{lemma:all_coalitions}, as we show in Section~\ref{SEC:more_than_pairs}.

For the upper bound, we take a coalition $R$ that is sampled uniformly from all $M_r$ possible coalitions, and bound the probability that it contains a deviating pair. Recall that from the proof of Theorem~\ref{th:convex}(a), a pair $k,j$ can deviate only if they are at most $\sqrt{s-j}\leq \sqrt s$ slots apart.

A coalition of size $r$ contains ${r \choose 2} = O(r^2)$ pairs, and each such pair has a probability of at most $\frac{2\sqrt{s}}{s} = O\(\frac{1}{\sqrt{s+r}}\) = O\(\frac{1}{\sqrt s}\)$. From the union bound we get that the probability that a random coalition $R$ contains \emph{any} deviating pair is at most $O\(\frac{r^2}{\sqrt s}\)$.
\end{proof}

\begin{proof}  [\proofholder of Proposition~\ref{th:LE_subsets}, lower bound]
If $r\geq s/2$, then $R$ contains a pair of neighbors and therefore surely has a deviation. Similarly, if $r =\omega(\sqrt s)$, then $R$ contains a neighbor pair with high probability. Assume therefore that $r$ is relatively small w.r.t. $s$, say  $r< s^{2/3}$. Note that for all $t\leq r$, ${s \choose t} = {s \choose t-1}\frac{s-t}{t} \geq  {s \choose t-1}s^{1/3}$. By induction, ${s \choose r} \geq {s \choose t} s^{1/3\cdot (r-t)}$.

Let $c<1$ be a constant, $d=\sqrt c$.

\begin{lemma} For a sufficiently large $s$, ${s \choose r} > d \cdot M_r$.
\label{lemma:M_sum}
\end{lemma}

\begin{proof}
Consider the sum $M_{r-1} = \sum_{t=1}{r-1}{s \choose t}$. It holds that
$$ M_{r-1} \leq \sum_{t=1}^{r-1}{s \choose r}s^{1/3(t-r)} = {s \choose r}\sum_{t=1}^{r-1}(s^{1/3})^{-t} \leq 2s^{-1/3}{s \choose r}.$$
In particular, for a sufficiently large $s$, we have that $2s^{-1/3}< 1-d$, and thus $M_{r-1} <  (1-d) M_r.$
Recall that $M_r = M_{r-1} + {s \choose r}$, thus
$${s \choose r} = M_r - M_{r-1} > M_r - (1-d)M_r = d\cdot M_r.$$
\end{proof}

As we perform an asymptotic analysis, we indeed assume that $s$ is as large as required.

Let $q=d^{2/r}$. We consider coalitions of size $r$ in slots $1,2,\ldots,qs$ (i.e. coalitions of the first type only). We show that there is only a small fraction of the ${qs \choose r}$ coalitions \emph{do not} have a deviation. 

\begin{lemma} ${{qs}\choose r} \geq d \cdot {s \choose r}$.
\label{lemma:M_bound}
\end{lemma}
\begin{proof}
\begin{align}
q &= d^{2/r} = e^{2\ln(d) /r} > \(1+\frac{2\ln d}{r}\) \label{eq:q_val} \\
&\frac{{qs\choose r}}{{s \choose r}} = \frac{(qs)!(s-r)!}{s!(qs-r)!} = \prod_{t=1}^r \frac{qs-t}{s-t} \geq \(\frac{qs-r}{s-r}\)^r > \(\frac{s+\frac{2\ln d}{r}s-r}{s-r}\)^r \tag{from \eqref{eq:q_val}} \\
&= \(1+\frac{2s\ln d}{r(s-r)}\)^r > \(1+\frac{\ln d}{r-1}\)^r \geq \exp\(\frac{\ln(d)r}{r}\) = \exp(\ln d) = d. \notag
\end{align}
\end{proof}

Also, from Equation~\eqref{eq:q_val},
\labeq{sqrt_q}
{\sqrt{1-q} \geq \sqrt {1-\(1+\frac{2\ln d}{r}\)} = \sqrt \frac{-2\ln(d)}{r} > d'\frac{1}{\sqrt r},}
where $d'>0$ is some constant independent of $r$ and $s$.

We construct our coalition iteratively, lower bounding in every iteration the probability that a deviating pair is formed. Since all bidders are in slots $\leq qs$, it is sufficient for the first pair $k,j$ to deviate if they are at most $\sqrt{s-j} \geq \sqrt{s-qs} = \sqrt{s(1-q)}$ slots apart. If the first pair are too far away, the third selected bidder has a double chance to have a deviation (with at least one of them). If this fails, the fourth bidder can be in the proximity of either of the first three, and so on.

Denote by $E_t$ the event that the bidder selected in iteration $t$ has a deviation with one of the previous bidders.
Suppose that none of the $t-1$ previous bidders has a deviation. The new bidder $t$ has $qs-(t-1)$ available slots. There are \emph{at least} $(t-1)\sqrt{s(1-q)}$ slot that are in the proximity of previous bidders, since there is a ``dangerous'' interval of size (at least) $\sqrt{s(1-q)}$ around each bidder, and these intervals are distinct (otherwise there is a deviating pair). Formally, this can be written as
\labeq{E_bound}
{Pr(E_t| \forall t'< t, \neg E_{t'}) \geq \frac{(t-1)\sqrt {s(1-q)}}{qs-t+1}.}

We have that for a random coalition $R$ drawn from $1,2,\ldots,qs$, the probability that $R$ does \emph{not} contain a deviating pair, is
\begin{align*}
Pr(\neg E_t &\text{ for all } t=2,3,\ldots,r) = Pr(\neg E_2)Pr(\neg E_3 | \neg E_2)\cdots Pr(\neg E_r| \forall t'< r, \neg E_{t'}) \\
&\leq \(1-\frac{2\sqrt {s(1-q)}}{qs-1}\)\cdots\(1-\frac{(r-1)\sqrt {s(1-q)}}{qs-r+1}\) \tag{from \eqref{eq:E_bound}}\\
&< \prod_{t=1}^{r-1}\!\!\(1-\frac{t\sqrt {s(1-q)}}{s}\) \leq \prod_{t=\floor{r/2}}^{r-1}\!\!\(1-\frac{t\sqrt {s(1-q)}}{s}\)\\
&\leq \prod_{t=\floor{r/2}}^{r-1}\!\(1-\floor{\frac{r}{2}}\frac{\sqrt {s(1-q)}}{s}\) \leq \(1-\floor{\frac{r}{2}}\frac{\sqrt {(1-q)}}{s}\)^{\ceil{\frac{r}{2}}-1} \\
&\leq\!\(\!1-\frac{(r-1)\sqrt {1-q}}{2\sqrt s}\)^{\frac{r-2}{2}} \! \leq\! \(\!1-d'\frac{r-1}{2\sqrt s\sqrt r}\)^{\frac{r-2}{2}} \tag{from \eqref{eq:sqrt_q}} \\
& \leq \exp\(-d'\frac{(r-1)(r-2)}{4\sqrt s\sqrt{r}}\) = \exp\(-\Omega\(\frac{r\sqrt r}{\sqrt s}\)\)
\end{align*}
Thus there are at least ${qs \choose r} \(1-\exp\(-\Omega\(\frac{r\sqrt r}{\sqrt s}\)\)\)$ coalitions of size $r$ with deviations.
Finally, we get from Lemmas~\ref{lemma:M_bound} and \ref{lemma:M_sum} that ${qs \choose r} \geq  d {s \choose r} \geq d^2 M_r = c M_r$, thus
\begin{align*}
\mc D_r(GSP,LE) \geq c \cdot M_r \cdot \(1-\exp\(-\Omega\(\frac{r\sqrt r}{\sqrt s}\)\)\) \\
\end{align*}
as required.
\end{proof}

\apxsection{Proofs of Section~\ref{SEC:RESERVE}}
\label{apx:reserve}
VCG with reserve price $c$ is typically defined as follows:  remove bidders whose value is below $c$. Now run VCG on remaining bidders. 

In our definition, we said that each remaining bidder pays the maximum between her original VCG payment and $c$.
\begin{proposition}
The two definitions are equivalent. 
\end{proposition}
\begin{proof}
Let $p_i$ denote the original payment of agent $i$ in VCG \emph{without} a reserve price. $p'_i$ is the payment with reserve price according to the first definition, and $p''_i$ is the payment according to the second definition. That is, $p''_i = \max(c,p_i)$ if $v_i\geq c$ and 0 otherwise.

Let $\alpha_i = \frac{x_i}{x_{i-1}}$. According to Varian(?), $p_i = b_{i+1}$, where $b_i$ is recursively defined as follows. $b_{s+1}=v_{s+1}$, and
$$p_{i-1} = b_i = \alpha_i  v_i + (1-\alpha_i) b_{i+1}.$$

Let $j$ be the index of the lowest surviving bidder. Clearly if $j\geq s+1$ then both auctions coincide with the original VCG auction, as the reserve price is not used at all. Therefore suppose $j\leq s$.

We now turn to compute $p'_i$ in the same way. Suppose we add a positive term $\delta$ to all valuations. Then clearly all payments will also increase by $\delta$.

Since all values (after removing the low bidders) are above $c$, we can decrease all $v_i$ by $c$, to $v'_i = v_i-c$, and add $c$ to the final payment. That is, 
$\vec p' = \vec p^*+c$, where $\vec p^*$ are the VCG payments for valuations $\vec v'$. We claim that $p^*_i = p_i-c$. The base case of the induction is $p^*_j = b^*_{j+1} = 0$ (since there are at most $s$ bidders). The next bidder pays $p^*_{j-1} = b^*_{j} = \alpha_j v'_j = \alpha_j (v_j - c)$. 
\end{proof}

%